# Shape of a slowly rotating star measured by asteroseismology

Laurent Gizon[1,2,3,4,*], Takashi Sekii[3], Masao Takata[5], Donald W. Kurtz[6], Hiromoto Shibahashi[5], Michael Bazot[4], Othman Benomar[4,5], Aaron C. Birch[1], Katepalli R. Sreenivasan[7,4]

[1]Max-Planck-Institut für Sonnensystemforschung, 37077 Göttingen, Germany. [2]Institut für Astrophysik, Georg-August-Universität Göttingen, 37077 Göttingen, Germany. [3]National Astronomical Observatory of Japan, Mitaka, Tokyo 181-8588, Japan. [4]Center for Space Science, NYUAD Institute, New York University Abu Dhabi, PO Box 129188 Abu Dhabi, UAE. [5]Department of Astronomy, The University of Tokyo, Bunkyo-ku, Tokyo 113-0033, Japan. [6]Jeremiah Horrocks Institute, University of Central Lancashire, Preston PR1 2HE, UK. [7]New York University, NY 10012, USA.
[*]Corresponding author. E-mail: gizon@mps.mpg.de.

**Stars are not perfectly spherically symmetric. They are deformed by rotation and magnetic fields. Until now the study of stellar shapes has only been possible with optical interferometry for a few of the fastest-rotating nearby stars. Here we report an asteroseismic measurement, with much better precision than interferometry, of the asphericity of an A-type star with a rotation period of 100 days. Using the fact that different modes of oscillation probe different stellar latitudes, we infer a tiny but significant flattening of the star's shape of $\Delta R/R = (1.8 \pm 0.6) \times 10^{-6}$. For a stellar radius $R$ that is 2.24 times the solar radius, the difference in radius between the equator and the poles is $\Delta R = 3 \pm 1$ km. Because the observed $\Delta R/R$ is only one third of the expected rotational oblateness, we conjecture the presence of a weak magnetic field on a star that does not possess an extended convective envelope. This calls to question the origin of the magnetic field.**

## INTRODUCTION

According to Clairaut's theorem, slowly rotating stars are oblate spheroids (*1*, *2*). Other factors may affect the shapes of stars, such as magnetic fields, thermal asphericities, large-scale flows, or strong stellar winds. Global poloidal magnetic fields tend to make stars oblate, while toroidal magnetic fields tend to make them prolate (*3*, *4*). The tidal interaction of a star with a close stellar companion or a giant planet is yet another cause of stellar deformation (*5*). Thus measuring the asphericity of stars can place constraints on a wide range of phenomena beyond the standard model of stellar structure and evolution. Direct imaging of the deformed shapes of nearby stars requires a resolution better than a milliarcsecond. The elongated projected shape of the rapidly rotating A star Altair has been observed with infrared interferometry to have a flattening $\Delta R/R = 0.14 \pm 0.03$ (*6*, *7*). Vega, another rapidly rotating A star, has an apparent deformation that is too small to be measured as it is seen almost pole-on (*8*). In this paper we present with unprecedented precision the first measurement of stellar asphericity by means of asteroseismology (*9*), for the star KIC 11145123, which has an equatorial velocity two orders of magnitude smaller than either Altair or Vega. This work is motivated by helioseismology's ability to probe the Sun's asphericities and their temporal variations with the eleven-year solar magnetic cycle (*10*, *11*).

**RESULTS**

The star KIC 11145123 belongs to the class of hybrid pulsators (*12*). It oscillates both in a high-frequency band (15-25 day$^{-1}$) and in a low-frequency band (below 5 day$^{-1}$). The observed modes of oscillation are acoustic (p), gravity (g), and mixed (p and g) modes. Modes with dominant p-mode character are seen in the high-frequency band. These modes oscillate throughout most of the star, with larger oscillation amplitudes near the surface. They are labelled with the radial order, *n*, which counts the number of nodes in the radial direction with a positive sign for nodes in the p-mode cavity and a negative sign for nodes in the g-mode cavity. Most known hybrid pulsators, including KIC 11145123, belong to the γ Doradus-δ Scuti class (*13*). Oscillations in these stars are likely to be excited by the opacity (p and mixed modes) and the convective-blocking (g modes) mechanisms.

Oscillations of KIC 11145123 were observed in intensity over a period of $T = 3.94$ years by Kepler (*14*). Because the oscillations are purely harmonic, random errors in the inferred mode frequencies scale as $T^{-3/2}$ times the noise-to-signal ratio of the periodic oscillations, and thus the mode frequencies can be determined with astounding precision (*15*). In the p-mode frequency band, Kurtz et al. (*12*) report frequency errors between $5 \times 10^{-7}$ day$^{-1}$ and $10^{-4}$ day$^{-1}$. The stellar model that best explains the observed mode frequencies implies that KIC 11145123 is a terminal-age main sequence (TAMS) A star. It has a small convective core ($r < 0.04\ R$) in which the fraction of hydrogen content is less than 5%. Outside this convective core, energy is transported by radiation up to the surface layers of the star. In the top few thousand kilometers there are very thin convective layers associated with the ionization of helium and hydrogen. See Table 1 for a summary of the basic stellar parameters.

In spherically symmetric stars, the eigenfunctions of stellar oscillations are proportional to spherical harmonics $Y_l^m(\theta, \varphi)$ of degree $l$ and azimuthal order $m = -l, -l + 1, \ldots l$, where $\theta$ is the co-latitude and $\varphi$ is the longitude. Internal rotation and departures from spherical symmetry lift the $(2l + 1)$-fold degeneracy in $m$ of the mode frequencies, $\nu_{nlm}$. The antisymmetric component of the frequency splittings in a multiplet, $\delta\nu_{nlm} = (\nu_{nlm} - \nu_{nl,-m})/2$, is a weighted average over the stellar volume of the stellar angular velocity (*16*). KIC 11145123 is one of a very few stars in which these rotational splittings have unambiguously been detected both in the p-mode and g-mode bands. The observed frequency splittings imply an internal rotation period of more than 105 days and a surface rotation period of less than 99 days, showing that the star rotates a little more quickly at the surface than in the core (*12*). Internal angular momentum transfer or external accretion mechanisms have spun up the atmosphere, a result of theoretical interest (*17*).

Stellar asphericity is measured through the symmetric component of the splittings (*9*):
$$s_{nlm} = (\nu_{nlm} + \nu_{nl,-m})/2 - \nu_{nl0}, \qquad \text{for } 1 \leq m \leq l. \tag{1}$$
This differential measurement exploits the different sensitivity in latitude of modes with different values of $|m|$ at fixed $l$ and $n$. As shown in Fig. 1, modes with larger values of $|m|$ are confined to lower latitudes. For p modes, the mean frequencies $\bar{\nu}_{nlm} = (\nu_{nlm} + \nu_{nl,-m})/2$ are not sensitive to rotation at first order and inform us about the inverse acoustic stellar radius at specific latitudes, with increased sensitivity at lower latitudes for larger $|m|$. For a spherical star, $\bar{\nu}_{nlm} = \nu_{nl0}$ and $s_{nlm} = 0$ for all $m$. Latitudinal variations in stellar shape or wave speed will cause a non-zero $s_{nlm}$. The $s_{nlm}$ are positive for prolate spheroids and negative for oblate spheroids. Latitudinal variations in the wave speed may result from variations in a magnetic field or chemical composition.

In the p-mode frequency band of KIC 11145123, five multiplets have been identified (*12*) and assigned values of $(n, l, m)$ by comparison with the best-fit seismic model. These are two dipole ($l = 1$) and three quadrupole ($l = 2$) multiplets, for which all $2l + 1$ azimuthal modes are identified. The measured values of $s_{nlm}$ are tabulated in Table 2. Among these, the quadrupole multiplet near 23.5 day$^{-1}$ does not provide frequencies with sufficient precision to affect the results of this paper. The $s_{nlm}$ are plotted in Fig. 2 for the four multiplets of interest. The average of all values, $\langle s_{nlm} \rangle = (-1.4 \pm 0.5) \times 10^{-5}$ day$^{-1}$, is negative (three standard deviations away from zero): the star is oblate.

As mentioned in the introduction several physical mechanisms can make a star aspherical to stellar oscillations. One mechanism that must be present is rotational oblateness, which is relatively easy to compute when rotation is slow. The centrifugal force distorts the equilibrium structure of a rotating star. The corresponding perturbation to the mode frequencies scales as the ratio of centrifugal to gravitational forces,

$$\varepsilon = \Omega^2 R^3/(GM), \qquad (2)$$

where $\Omega$ is the star's surface angular velocity, $R$ and $M$ are the radius and mass of the star, and $G$ is the universal constant of gravity. Using $\Omega/2\pi = 0.01$ day$^{-1}$ for KIC 11145123, we have $\varepsilon = 1.34 \times 10^{-6}$ $(R/R_\odot)^3 (M/M_\odot)^{-1}$, where $R_\odot$ and $M_\odot$ are reference solar values (*18*). The mass and radius of the star are not known to the same level of confidence as rotation. Our best-fit seismic model (*12*) has a metallicity $Z = 0.01$, a mass of 1.46 $M_\odot$, and a radius of 2.24 $R_\odot$. For this stellar model, the ratio of the centrifugal to gravitational forces becomes

$$\varepsilon = 1.0 \times 10^{-5}. \qquad (3)$$

This is a very small number, but not small compared to the relative errors of the most precise frequencies in the p-mode range from Table 2. Note that $\varepsilon$ is roughly half the solar value ($\varepsilon_\odot = 1.8 \times 10^{-5}$).

For slow rotators, rotational oblateness is described by a quadrupole distortion of the stellar structure. To leading order, the contribution of rotational oblateness to $s_{nlm}$ can be written as (*19*, *16*)

$$s^{rot}_{nlm} = -\varepsilon\, m^3\, v_{nl}\, \Delta_{nl}\, (2l-1)^{-1}\, (2l+3)^{-1}, \qquad (4)$$

where the dependence on $m$ and $l$ is due to the latitudinal sensitivity of the modes of oscillation (Fig. 1). The amplitude of the effect is proportional to the degenerate mode frequency $v_{nl}$ of the non-rotating reference model and to the numbers $\Delta_{nl}$, which are mode-weighted radial averages of the stellar distortion (see Table 2 and Materials and Methods section). The numerical values of $s^{rot}_{nlm}$ are listed in Table 2 and are overplotted in Fig. 2 for the available modes. We find that the theoretical $s^{rot}_{nlm}$ are of the same sign and same order of magnitude as the measured $s_{nlm}$. As illustrated in Fig. 3, a good representation of the measurements is

$$s_{nlm} = (0.35 \pm 0.12)\, s^{rot}_{nlm}. \qquad (5)$$

Hence the star is more round than rotational oblateness would imply. Equation 5 also implies that the modes of oscillation see a quadrupole distortion of the shape of the star. The amplitude of the distortion is smaller than would be expected from rotation alone: an additional physical ingredient is needed.

**DISCUSSION**

The flattening of the stellar surface due to rotation alone would be $(\Delta R/R)^{rot} = \varepsilon/2 = 5 \times 10^{-6}$, where $\Delta R$ is the difference between the equatorial and polar radii. Here, the effective flattening of the stellar surface implied by the seismic measurements (Eq. 5) is only

$$\Delta R/R = (1.8 \pm 0.6) \times 10^{-6}. \tag{6}$$

To our knowledge, KIC 411145123 is the most spherical natural object ever measured, more spherical than the quiet Sun (*32*).

Using $R = 2.24\ R_\odot = 1.56 \times 10^6$ km, we have $\Delta R = 2.7 \pm 0.9$ km. This is an astonishing illustration of the remarkable precision of this asteroseismic diagnostic and a direct consequence of the very long lifetime of the oscillations under study. There is however a limitation in accuracy due, mainly, to the uncertainty in the radius of the stellar model. We may incorporate the uncertainty in the stellar radius in the error for $\Delta R$; the conservative assumption of a systematic error of one solar radius implies a combined error of 1.5 km on $\Delta R$. We stress that the uncertainty in the stellar radius is a systematic error that does not change the 3-σ significance level of the result; it only changes the absolute value of $\Delta R$.

Guided by the well-established results of helioseismology (*11*, *20*), we suggest here that a weak surface magnetic field (much weaker than the Sun's at solar maximum, see Materials and Methods) is a possible explanation for the reduced oblateness of KIC 11145123: waves propagate faster in magnetized regions, so surface magnetic fields at low latitudes will make a star appear less oblate to acoustic waves. We note that observations of photometric variability have led to the speculation that a large fraction of A stars have starspots (*21*). Yet the origin of magnetic fields in stars without deep convective envelopes is a matter of debate (*22*). Dynamo action may take place in the core or in the very thin convective layers near the surface, or the magnetic field may have a fossil origin.

Other than a magnetic field, there are few alternative explanations for the reduced oblateness. At this level of precision, the physics of stellar oscillations may need to be studied in more detail. In particular, it is not quite excluded that nonlinear (amplitude) effects could play a role; this should be investigated further. On the other hand, nonadiabatic effects are spherically symmetric and will not affect $s_{nlm}$ to leading order.

Nearly all slowly-rotating A stars have overabundances of certain metal elements (*23*). That KIC 11145123 is not a chemically peculiar star (Am or Ap) is surprising, hence the speculation in Ref. (*12*) about possible blue straggler mass transfer. An enhancement or deficiency of metals in the atmosphere would only affect seismic asphericity if the abundances were non-uniformly distributed in latitude. That could happen in magnetic Ap stars, but a Subaru high-resolution spectrum does not show Ap abundances and shows a metal deficiency of 0.7 dex. While we cannot rule out a latitudinal gradient in chemical composition, this explanation is more involved than a weak magnetic field.

Since stars more massive than the Sun are more likely to harbor giant planets (*24*), one may also ask if KIC 11145123 could be deformed by tidal interaction. In the linear regime, only the equilibrium tidal deformation should be considered. However it is smaller than the rotational deformation by a factor proportional to the ratio of the mass of the planet to the mass of the star (*19*, *25*); thus it is negligible for Jupiter mass planets. Furthermore, a planetary companion (or a stellar companion) in the equatorial plane of the star would make the star look more oblate to the acoustic modes, not less oblate as required by the observations.

This work is a first step in the study of stellar shapes through asteroseismology. The method demonstrated here will be applied to other stars, including more rapidly rotating stars and stars with stronger magnetic fields, where deformations will be greater. Thanks to the unprecedented high precision and long time-span of the *Kepler* observations, an important field of theoretical astrophysics is now also observational.

**MATERIALS AND METHODS**

**Mode frequency measurements**

The frequencies of the modes of oscillation were measured using *Kepler* light curves for quarters Q0-16, spanning a total of 51 months of data. The mode frequencies were determined by nonlinear least squares in the time domain assuming Gaussian uncorrelated errors; see Ref. (*12*) for a full description of the data reduction. We tested a new frequency solution on *Kepler* quarters Q0-17 and found that there were no significant changes compared to the published Q0-16 analysis. So that our work could be easily tested and reproduced by others, we chose to use the published data (*12*).

The frequency errors were determined using an estimate of the variance around each mode frequency. Because many nearby frequency peaks may contribute to this variance, the frequency errors are conservative. Had all significant peaks been removed from the amplitude spectrum, the error estimates would have been smaller.

The *Kepler* data are averaged over consecutive 29.4-min time intervals, i.e. a significant fraction of the shortest p-mode periods (~ 1 hour). This effect reduces the observed amplitudes of the modes but does not affect the mode frequencies. The only effect is a reduced signal-to-noise ratio (compared to shorter integration times); this ratio was taken into account in the estimation of the errors on mode frequencies.

**Effect of centrifugal distortion on mode frequencies**

The effect of the centrifugal force on mode frequencies can be evaluated using second-order perturbation theory, either in the spherical geometry of the reference model (*26*) or in the distorted geometry of the oblate spheroid (*19*, *27*). It consists of several terms, which account for geometrical distortion, change in wave speed, and first-order perturbation to the mode eigenfunctions. In the p-mode frequency range, for which $[\Omega/(2\pi\nu_{nl})]^2 \ll \varepsilon$, the effect of rotation on mode frequencies is well approximated by (*16*)

$$\nu_{nlm} \approx \nu_{nl} + m\Omega/(2\pi) + \varepsilon\, \nu_{nl}\, \Delta_{ln}\, Q_{2lm}\, , \qquad (7)$$

where $\nu_{nl}$ is the mode frequency in the non-rotating reference stellar model,

$$Q_{2lm} = \int_0^\pi d\theta\, P_2(\cos\theta)\, E_{lm}(\theta) = [l(l+1) - 3m^2]\,(2l-1)^{-1}\,(2l+3)^{-1} \qquad (8)$$

is the latitudinal average of quadrupole distortion weighted by latitudinal mode energy density $E_{lm}$ (see Fig. 1), and the dimensionless number

$$\Delta_{nl} = 4/3 \int_0^R dr\, (r/R)^3\, \xi_{nl}^2(r)\, \rho r^2 \qquad (9)$$

is the radial average of the centrifugal distortion weighted by mode energy density (Fig. S1), which depends on the (normalized) radial mode displacement $\xi_{nl}$. For modes with pure p-mode character, $\Delta_{nl} \approx 0.7$. For the modes under consideration here, $\Delta_{nl}$ is in the range 0.2-0.7 (Table 2), where the smaller values are for the mixed modes (Fig. S2). The above expression, Eq. 9, assumes rigid body rotation and neglects the perturbation to the gravitational potential related to the star's quadrupole moment; both approximations are at the level of a few percent (*28*) and are thus acceptable for the purpose of estimating the contribution of rotational oblateness to $s_{nlm}$. Combining the definition of $s_{nlm}$ (Eq. 1) and Eq. 7, we obtain Eq. 4.

**Alternative stellar model**

We note that the effective temperature of the best-fit seismic model is not consistent with the photometric value (Table 1). This prompted Kurtz et al. (*12*) to consider an alternative stellar model with $M = 2.05$ M$_\odot$ and $R = 2.82$ R$_\odot$, whose effective temperature is within error bars. This model is however a worse fit to the p-mode frequencies, making mode identification more difficult. In particular, in Table 2, only the $l = 2$ mode at 16.7 day$^{-1}$ and the $l = 1$ mode at 18.4 day$^{-1}$ can be identified. For the alternative model we have $\varepsilon = 1.5 \times 10^{-5}$, which is 50% larger than for the best-fit seismic model. Should this alternative stellar model be preferred, the estimates of $s_{nlm}^{\rm rot}$ should be scaled appropriately so that $s_{nlm} = (0.23 \pm 0.08)\, s_{nlm}^{\rm rot}$ and $\Delta R = 3.4 \pm$ 1.1 km. We note that this stellar model and the best-fit seismic model referred to in the main text were obtained from a stellar evolutionary code that does not include rotation or magnetic fields.

**Helioseismology and upper limit on magnetic field**

In helioseismology, azimuthal mode frequencies in a multiplet are traditionally expanded as $a$-coefficients on a basis of Clebsh-Gordan polynomials (*29*). The odd $a$-coefficients, $a_{2k+1}$, are measures of differential rotation, while the even $a$-coefficients, $a_{2k}$, are measures of asphericity ($k = 0, 1, 2…$). The $s_{nlm}$ are related to the even $a$-coefficients. In particular, for dipole modes, $s_{n11} = 3a_2$ informs us about the $P_2$ component of distortion. The effect of solar rotational oblateness is too small ($a_2 \sim -10$ nHz for $l \leq 2$ modes) to be measured on individual multiplets (*11*). However solar asphericity measurements are possible by averaging over sets of intermediate degree modes ($l < 150$). When the Sun's magnetic activity is very low, $a_2$ coefficients are negative but of smaller magnitude than implied by rotational oblateness (*11*). During maxima of solar activity, the solar $a_2$ coefficients become positive ($a_2 \sim 100$ nHz for $l < 5$ modes) (*20*) and the Sun appears prolate to the acoustic modes: they sense magnetic activity at mid to low latitudes ($< 40°$). At the solar surface, the quadrupole components of the solar magnetic field vary by less than 10 Gauss with the sunspot cycle (*30*). Baldner et al. (*20*) used measurements of solar $a_2$ coefficients to infer toroidal and poloidal magnetic field components below the solar surface, at the level of a few hundred Gauss.

In light of the solar observations, a possible explanation for KIC 11145123's reduced oblateness is a magnetic perturbation (*9*). Let us consider the dipole mode at 18.3 day$^{-1}$, for which $\Delta a_2 = a_2 - a_2^{rot} = 1.8 \times 10^{-5}$ day$^{-1}$ = 0.2 nHz. By comparison with the solar observations and given that $\Delta a_2$ is expected to scale like the square of the magnetic field, we infer that a much smaller level of magnetism than the Sun's would be needed to explain the observations. However, it is difficult to be more specific as $\Delta a_2$ depends sensitively on the geometrical configuration of the magnetic field (*20*).

**SUPPLEMENTARY MATERIALS**

Supplementary material for this article is available at http://advances.sciencemag.org/
Fig. S1. Kinetic energy density of the modes of oscillation.
Fig. S2. Radial dependence of centrifugal distortion.

**Acknowledgments:** Useful discussions with Hideyuki Saio and Todd Hoeksema are gratefully acknowledged. **Funding:** L.G. acknowledges generous support through a visiting professorship at the National Astronomical Observatory of Japan, Mitaka, Tokyo. L.G., M.B., O.B. and K.R.S. acknowledge research funding from the NYU Abu Dhabi Institute under grant G1502. M.T. acknowledges financial support from JSPS KAKENHI grant number 26400219. **Author contributions:** L.G., T.S., and M.T. designed the research. D.W.K. performed multiple tests to validate mode frequencies and associated errors. L.G. drafted the paper and all authors contributed to the final manuscript. **Competing interests:** The authors declare that they have no competing interests. **Data and materials availability:** The mode eigenfunctions may be requested from the authors.

**FIGURES AND TABLES**

**Table 1. Parameters of star KIC 11145123 and best-fit seismic model.**

|  | Photometry (*12, 31*) | Seismology (*12*) |
|---|---|---|
| Spectral type | A | late main sequence |
| *Kepler* visual magnitude | 13 | |
| Effective temperature, $T_{\text{eff}}$ | 8050 ± 200 K | 7032 K |
| Surface gravity, $\log g$ (cgs) | 4.0 ± 0.2 | 3.904 |
| Surface rotation rate | | $\Omega/2\pi = 0.01$ day$^{-1}$ |
| Mass | | $M = 1.46\ M_\odot$ |
| Radius | | $R = 2.24\ R_\odot$ |
| Initial abundances | | $(X, Z) = (0.65, 0.010)$ |
| Radius of convective core | | $0.04\ R$ |
| Hydrogen core abundance | | $X_c = 0.033$ |

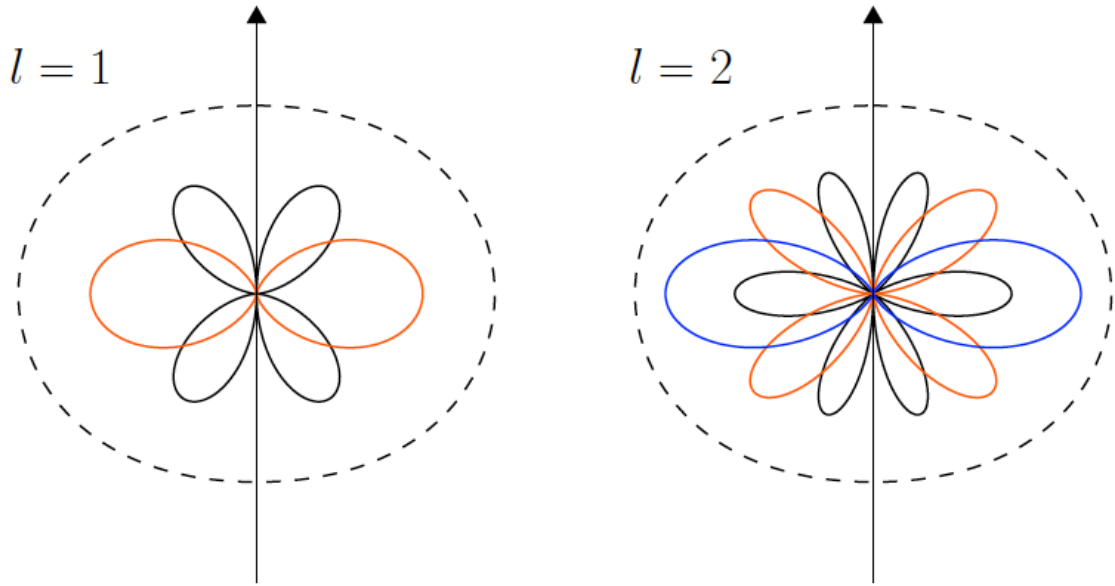

**Fig. 1. Latitudinal dependence of mode kinetic energy density.** For the dipole ($l = 1$, left panel) and quadrupole ($l = 2$, right panel) modes of oscillation. The arrow points along the stellar rotation axis. Scalar eigenfunctions of stellar oscillation are proportional to $P_l^m(\cos\theta)\, e^{im\varphi}$, where the $P_l^m$ are associated Legendre functions. Shown are polar plots of the kinetic energy density $E_{lm}(\theta) = c_{lm}\,[P_l^m(\cos\theta)]^2\,\sin\theta$, where $\theta$ is co-latitude, for the modes with azimuthal orders $m = 0$ (black), $m = 1$ (red), and $m = 2$ (blue). The constants of normalization, $c_{lm}$, are such that $\int_0^\pi d\theta\, E_{lm}(\theta) = 1$ for each $(l, m)$. For dipole modes, we see that $E_{10}$ is maximum at latitude $\lambda = \pi - \theta = \pm 63°$ and $E_{11}$ is maximum at the equator. For quadrupole modes, $E_{20}$ peaks at $\lambda = 0°$ and $\pm 59°$, $E_{21}$ is maximum at $\lambda = \pm 39°$, and $E_{22}$ is maximum at the equator. For reference, the dashed curves show a highly distorted (oblate) stellar shape of the form $r(\theta) = 1 - 0.15\, P_2(\cos\theta)$, where $P_2(x) = (3x^2 - 1)/2$ is the second-order Legendre polynomial.

**Table 2. Mode frequencies and frequency shifts.** Values of $s_{nlm}$, as defined by Eq. 1, are given for the dipole and quadrupole multiplets in the p-mode range of KIC 11145123. The frequency shifts expected from rotational oblateness, $s_{nlm}^{rot}$, are computed using Eq. 4. Mode identification is according to the best-fit stellar model with $R = 2.24$ R$_\odot$ and $M = 1.46$ M$_\odot$ *(12)*. Mode amplitudes are measured to a precision of 0.01 mmag.

| n | l | m | $\nu_{nlm} \pm \sigma_{nlm}$ [day$^{-1}$] | amplitude [mmag] | $s_{nlm}$ [$10^{-5}$ day$^{-1}$] | $\Delta_{nl}$ | $s_{nlm}^{rot}$ [$10^{-5}$ day$^{-1}$] | index |
|---|---|---|---|---|---|---|---|---|
| −1 | 2 | −2 | 16.725 8824 ± 0.000 0017 | 2.33 | | | | |
| | | −1 | 16.733 9455 ± 0.000 0186 | 0.21 | | | | |
| | | 0 | 16.742 0104 ± 0.000 0075 | 0.53 | | | | |
| | | 1 | 16.750 0755 ± 0.000 0110 | 0.36 | 0.1 ± 1.3 | 0.40 | −1.0 | 1 |
| | | 2 | 16.758 0083 ± 0.000 0504 | 0.08 | −6.4 ± 2.6 | 0.40 | −3.9 | 2 |
| 2 | 1 | −1 | 18.355 8305 ± 0.000 0029 | 1.36 | | | | |
| | | 0 | 18.366 0001 ± 0.000 0135 | 0.29 | | | | |
| | | 1 | 18.376 1210 ± 0.000 0034 | 1.13 | −2.3 ± 1.4 | 0.68 | −7.7 | 3 |
| 0 | 2 | −2 | 18.986 9603 ± 0.000 0078 | 0.50 | | | | |
| | | −1 | 18.996 7001 ± 0.000 0060 | 0.65 | | | | |
| | | 0 | 19.006 4482 ± 0.000 0056 | 0.69 | | | | |
| | | 1 | 19.016 1736 ± 0.000 0101 | 0.39 | −1.1 ± 0.8 | 0.39 | −1.1 | 4 |
| | | 2 | 19.025 9102 ± 0.000 0086 | 0.45 | −1.4 ± 0.8 | 0.39 | −4.3 | 5 |
| 3 | 1 | −1 | 21.993 3315 ± 0.000 0064 | 0.61 | | | | |
| | | 0 | 22.001 8915 ± 0.000 0286 | 0.14 | | | | |
| | | 1 | 22.010 4220 ± 0.000 0064 | 0.61 | −1.5 ± 2.9 | 0.68 | −9.2 | 6 |
| 2 | 2 | −2 | 23.545 5350 ± 0.000 0303 | 0.13 | | | | |
| | | −1 | 23.555 3428 ± 0.000 0582 | 0.07 | | | | |
| | | 0 | 23.565 1835 ± 0.000 0480 | 0.08 | | | | |
| | | 1 | 23.574 9885 ± 0.000 0607 | 0.06 | −1.7 ± 6.4 | 0.20 | −0.6 | 7 |
| | | 2 | 23.584 7898 ± 0.000 0163 | 0.24 | −2.2 ± 5.1 | 0.20 | −2.2 | 8 |

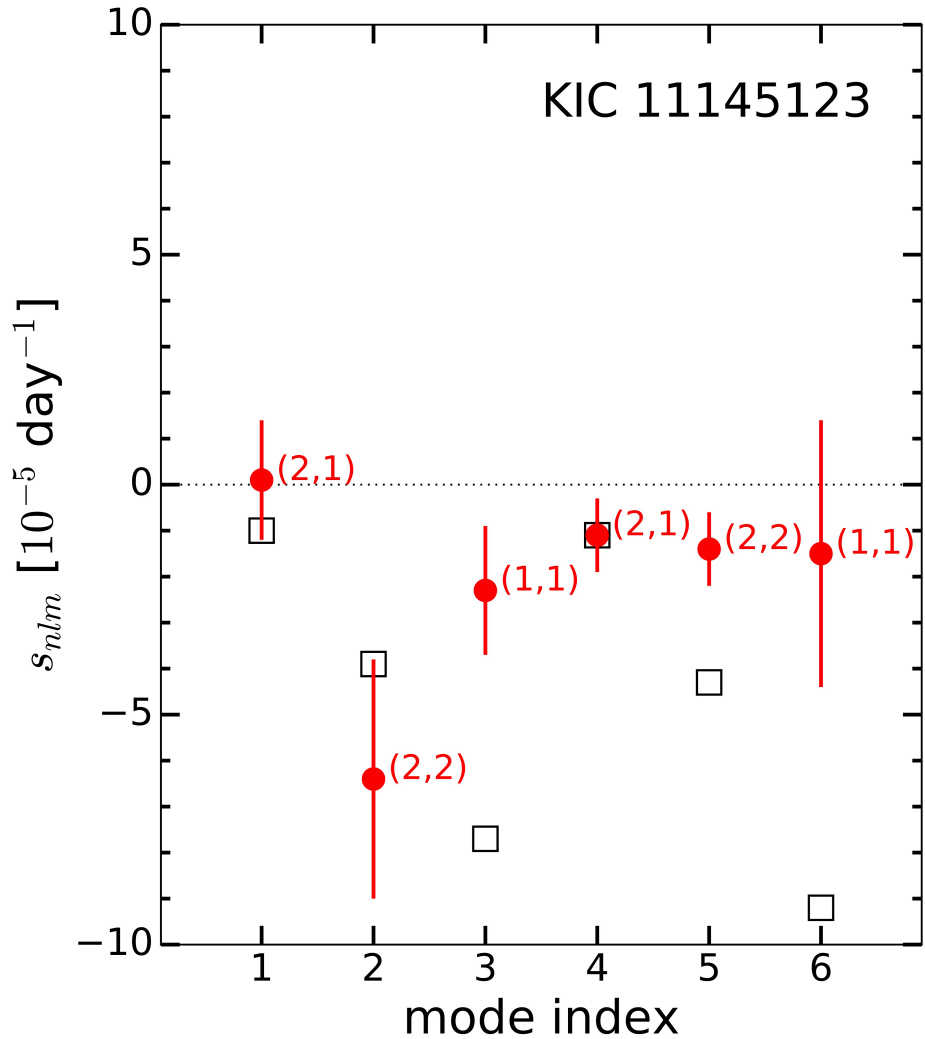

**Fig. 2. Symmetric component of observed frequency splittings $s_{nlm}$.** The observations are plotted as red circles with error bars. Each value is associated with the mode index given in the last column of Table 2. The data points are labeled by the values of ($l$, $m$). The theoretical values for rotational oblateness alone, $s_{nlm}^{\mathrm{rot}}$, are plotted as open squares. Notice that the last two values of $s_{nlm}$ from Table 2 are not plotted here as they are associated with errors that are too large to provide additional constraints.

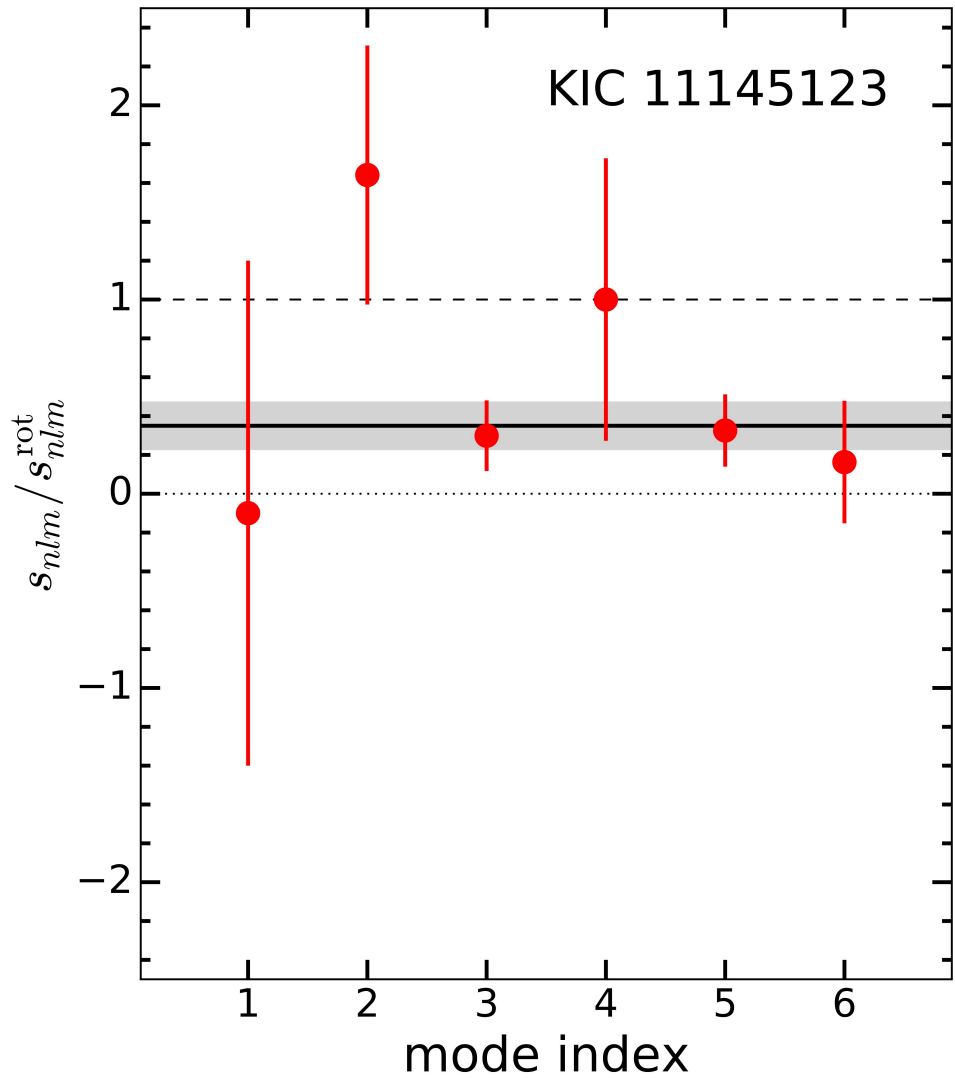

**Fig. 3. Ratios of observed $s_{nlm}$ to theoretical prediction for rotational oblateness $s_{nlm}^{\rm rot}$.** The horizontal solid line and the gray area indicate the average and the one-sigma bounds, $\langle s_{nlm}/s_{nlm}^{\rm rot} \rangle = 0.35 \pm 0.12$. Each value is associated with the mode index given in the last column of Table 2. The distributions of the data points and their errors are consistent with a single value for the ratio $s_{nlm}/s_{nlm}^{\rm rot}$.